# 3-D SPH Simulations of HD and MHD Jets

ELISABETE M. DE GOUVEIA DAL PINO & ADRIANO HOTH CERQUEIRA

*University of São Paulo, Instituto Astronômico e Geofísico,
Av. Miguel Stéfano, 4200, São Paulo, SP 04301-901, Brazil,
dalpino@astro1.iagusp.usp.br*

We present fully three-dimensional (3D) hydrodynamical (HD) and magnetohydrodynamical (MHD) simulations of supersonic, radiative cooling jets propagating into homogeneous and stratified environments based on the Smoothed Particle Hydrodynamics (SPH) technique.

1. INTRODUCTION

Supersonic, collimated outflows are a common phenomenum in the universe and span a large range of intensities and scaling, from the most powerful examples observed to emerge from the nuclei of active galaxies (or AGNs) to the jets associated to low-mass Young Stellar Objects (YSOs) within our own Galaxy. For a review of their observational properties see, e.g, Bridle & Perley (1984) and Biretta (1995) for AGN jets, and Raga (1993), Reipurth & Heathcote (1993) and Ray (1995) for YSO jets. AGN jets have typical velocities that approach c, lengths up to $\sim 1$ Mpc, and are believed to be *light*, i.e., less dense than their surroundings, while YSO jets have velocites $\sim 50$ to 400 km s$^{-1}$, lengths $\sim 1$ pc, and are *heavy*, with characteristic jet to ambient density ratio $\eta \simeq 1$ to 10. Despite their difference in scaling, both families of jets show strong morphological similarities, such as, a bow shock-like structure at the head where the jet impacts the ambient medium; approximately regularly spaced emission knots along the beam, which may be signatures of internal shock waves; and association with magnetic fields ($\sim 10^{-6}$ to $10^{-5}$ G) whose projected directions, in the case of the YSO jets, are mostly parallel to the jet axis.

Thanks to their proximity, YSO jets constitutes, perhaps, the best sources for cosmic jet investigation (Gouveia Dal Pino 1995) and in this work we examine their morphological properties with the aid of three-dimensional (3-D) numerical simulations of pure hydrodynamical (HD) and magnetohydrodynamical (MHD) jets.

2. THE NUMERICAL MODEL

The HD and MHD conservation equations are solved using the 3D, cartesian, smoothed particle hydrodynamics (SPH) technique described in Gouveia Dal Pino

& Benz (1993, hereafter GB93), (see also Gouveia Dal Pino & Benz 1994, hereafter GB94, Gouveia Dal Pino & Birkinshaw 1995, hereafter GB95). The SPH is a gridless, Lagrangean approach to fluid dynamics which employs particles that track and move with the fluid. The SPH particle properties are smoothed out in space by a spherically symmetric *kernel* function $W$ of width $h$. In this formalism, a variable $f$ of the system is given by:

$$< f(\vec{r}) >= \sum_k m_k \frac{f_k}{\rho_k} W(\vec{r} - r_k, h) \quad (1)$$

where $m_k$, $\rho_k$, and $r_k$, are the mass, the mass density, and the position of the particle $k$, respectively.

In the computational domain, the ambient gas is represented by a 3-D rectangular box filled with particles. A supersonic jet of radius $R_j$ is injected continually into the bottom of the box, which has dimensions up to $\sim 30R_j$ in the jet axis direction and $\sim 12R_j$ in the transverse directions.

The jet and the ambient gas are treated as a single ionized fluid with a ratio of specific heats $\gamma = 5/3$ and an ideal gas equation of state $p = u(\gamma - 1)\rho$, where $u$ is the internal energy per unit mass, and $p$ is the thermal pressure. The radiative cooling (due to collisional excitation and de-excitation, and recombination), which is relevant in YSO jets, is calculated using the cooling function evaluated for a gas of cosmic abundances cooling from $T = 10^6$ K to $T \approx 10^4$ K.

The evolution of the system is parameterized by the dimensionless numbers: i) $\eta = n_j/n_a$ (the ratio between the input jet and ambient number densities); (ii) $M_a = v_j/c_a$ (the initial ambient Mach number, where $c_a$ is the ambient sound speed); (iii) $k_p = p_j/p_a$ (the input pressure ratio which has been assumed to be initially equal to unity); and (iv) $q_{b_s}$ (the cooling length in the postshock gas behind the the bow shock, in units of the jet radius). The inclusion of the magnetic field in the MHD simulations adds a new parameter to the models $\beta$ (the ratio of the magnetic to the thermal pressure).

3. PURE HD SIMULATIONS

In previous 3D hydrodynamical modeling of continuous and intermittent jets performed with beams propagating into homogeneous ambient media (GB93, GB94, Chernin *et al.* 1994), we found the fundamental features observed in YSO jets (see also Raga *et al.* 1990, Hartigan & Raymond 1992, Kofman & Raga 1992, Stone & Norman 1993, Massaglia *et al.* 1995, Downes *et al.* 1995). The beam impacts supersonically the ambient material and develops a bow shock-like structure at the head, with a dense, cold shell composed of filaments and blobs formed from Rayleigh-Taylor and global thermal instabilities (GB93). The fragmented shell resembles the clumpy structure of observed HH objects at the heads of YSO jets (e.g., HH1, HH2, HH19, HH12).

The mechanism that produces the bright knots along the beam is still controversial, but the HD simulations indicate that the best candidate is probably intermittence in the ejection velocity of the jet (GB94; Raga *et al.* 1990, Hartigan & Raymond 1992, Kofman & Raga 1992, Stone & Norman 1993, Massaglia *et al.* 1995, Downes *et al.* 1995). Supersonic velocity variations (with periods close to the the dynamical time scale $\tau_{dy} = R_j/v_j$) quickly evolve to form a chain of regularly spaced radiative shocks along the jet with large proper motions. The knots widen and fade as they propagate downstream and eventually disappear (Fig. 1), so that this mechanism favors the formation of knots closer to the driving source in agreement with most of the observed systems. Longer variability periods (GB94) can also explain the multiple bow shocks observed in some jets (e.g., HH111, HH46/47 Reipurth & Heathcote 1993).

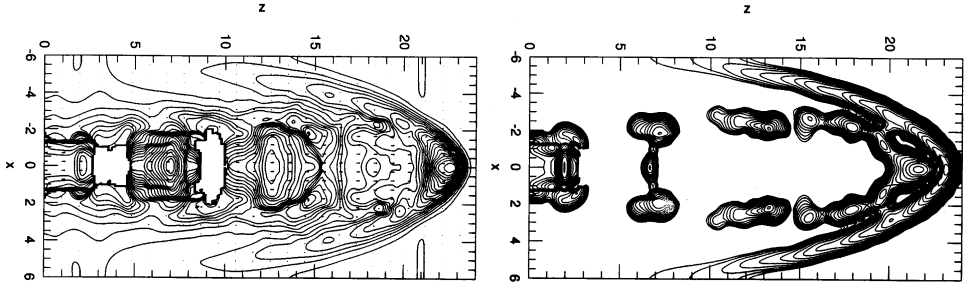

FIGURE 1. Mid-plane density (left) and radiative emissivity (right) contours of an intermittent, HD jet periodically turned on with a supersonic velocity $v_j = 150$ km s$^{-1}$ and periodically turned off with a subsonic velocity 15 km s$^{-1}$. The turning on and quiescent periods are given by 127 yrs. The initial parameters are $\eta = 10$, $n_a = 1000$ cm$^{-3}$, $R_j = 2 \times 10^{16}$ cm, and $M_a = 3$. The z and x coordinates are in units of $R_j$. The contour lines are separated by a factor of 1.2. The density ranges from $\simeq 0.01$ up to $42/n_a$, and the emissivity from $\simeq 0.01$ up to 5 (in code units). The evolution corresponds to $t \simeq 1386$ yrs.

More recently, motivated by the fact that the jets actually propagate into complex ambient media, we have performed HD simulations of jets propagating through more realistic environments (Gouveia Dal Pino et al. 1995; GB95). We have assumed isothermal ambient medium with a density (and pressure) along the jet axis given by $n_a(x) = n_a(x_o)[\alpha(x - x_o) + 1]^\nu$, where $n_a$ is the ambient number density, $x_o$ is the value of x at the jet inlet, $\alpha$ a parameter of the model and $\nu = \pm 5/3$ for positive and negative density gradients, respectively. In regions where the ambient gas has an increasing density (and pressure) gradient, we find that it compresses the cold, low-pressure cocoon that surrounds the beam and destroys the bow shock-like structure at the head. The compressing medium promotes the development of Kelvin-Helmholtz instabilities which over-confine the beam, cause some jet wiggling, and drive internal traveling shocks via pinching along the beam (Fig. 2). This could explain some observed YSO jets (e.g., Haro 6-5B jet; Mundt et al. 1991) which exhibit a wiggling structure with knot formation close to the jet head. In ambient regions of decreasing density (and pressure), we find that the flow widens and relaxes becoming very faint (Fig. 3). This could explain observed "invisible" jet sections like the gap between the parent source and the beam of Haro 6-5B. Due to their higher pressure cocoon, adiabatic jets are less affected by the stratification effects of a non-homogeneous ambient medium (GB95).

YSO jets appear to carry enough momentum ($\sim 6 \times 10^{-5} - 10^{-4} M_\odot$ yr$^{-1}$ km s$^{-1}$) to drive collimated, slower molecular outflows that are also associated to

YSOs (e.g., Henriksen 1995). Numerical HD simulations indicate that for YSO jets the transfer of momentum to the ambient medium occurs predominantly through the bow shocks and not by turbulent mixing at the lateral contact discontinuity (Chernin et al. 1994). (This latter mechanism is more relevant in low Mach number (Mj$\leq$ 3), light ($\eta \leq 3$) jets and then, more appropriate in AGN jets.) The molecular emission could thus be possibly identified with the environmental gas swept by the passage of internal bow shocks of a time-variable jet.

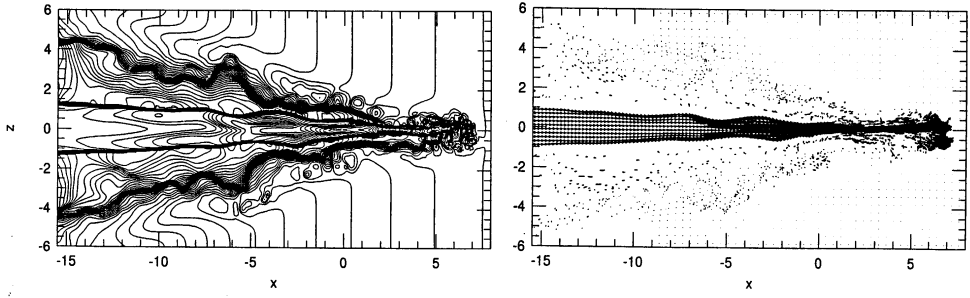

FIGURE 2. Mid-plane density contour (left) and velocity distribution (right) of a HD radiative cooling jet propagating into an ambient medium with positive density (and pressure) gradient ($\alpha = 0.5$, $\nu = 5/3$) at a time $t = 1.95$ (in units of 38.2 yr). The initial parameters are $\eta = 3$, $n_a = 200$ cm$^{-3}$, $R_j = 2 \times 10^{15}$ cm, $M_a = 24$, and $v_j = 398$ km s$^{-1}$. The contour lines are separated by a factor of 1.2. The density ranges from $\simeq 0.1$ up to $741/n_a$.

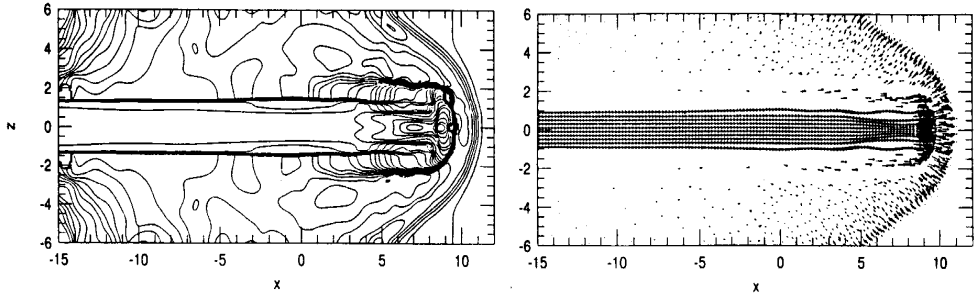

FIGURE 3. Mid-plane density contour (left) and velocity distribution (right) of a HD radiative cooling jet propagating into an ambient medium with negative density (and pressure) gradient ($\alpha = 0.5$, $\nu = -5/3$) at $t = 1.25$ (in units of 38.2 yr). The initial parameters are the same as in Fig. 2. The contour lines are separated by a factor of 1.3. The density ranges from $\simeq 0.02$ up to $26/n_a$.

4. MHD SIMULATIONS

While of fundamental importance in the production and initial collimation of jets, magnetic fields have been neglected in most of the analytical and numerical modeling of the structure of the YSO jets since the observational estimates of their intensity ($\sim 10^{-5}$ G) suggest that they are not dynamically important along the flow. However, they can become relevant when amplified by compression behind the shocks. We have then, incorporated the effects of magnetic fields in our 3D SPH simulations. As an example, Fig. 4 compares HD ($\beta = 0$) and MHD ($\beta = 1$) *radiative cooling* jets. (Previous 2D and 3D numerical work on MHD jets assumed an adiabatic approach; e.g., Clarke *et al.* 1986, Kössl *et al.* 1990, Todo *et al.* 1993.) We have assumed an initial helical force-free magnetic field with the toroidal ($B_\phi$) and longitudinal ($B_x$) components having the same profiles as in Todo *et al.* (1993). The initial $B_x(r)$ has a maximum strength $B \simeq 8 \times 10^{-5}$ G on the jet axis and decays with the radial distance r from the axis. The initial $B_\phi(r)$ has a maximum strength $\simeq 0.4B$ at $r = 3R_j$ and vanishes at $r = 0$. The larger collimation of the MHD jet is provided by the magnetic field which is amplified in the shocks at the head. The magnetic field also reduces, as expected, the growth of density in the shell at the head from $n_{sh}/n_a \simeq 240$ to $\simeq 65$. We note some beam pinching at x$\simeq -10, -7$, and 0, which arises from the developement of a radial $\mathbf{J}_x \times \mathbf{B}_\phi$ force at the contact discontinuity between the jet and the cocoon. This MHD instability drives internal oblique shock waves that may cause the formation of the observed emission knots associated with the YSO jets. We note also some kink helical instability (e.g., Chandrasekhar 1961) in development at the head of the MHD jet which is driven by $B_\phi$.

A more detailed description of our MHD results, with simulations covering a more extensive range of parameters and magnetic field geometries is presented elsewhere (Cerqueira & Gouveia Dal Pino 1996). The presence of radiative cooling inhibits the growth of the MHD instabilities due to the smaller effective (thermal plus magnetic) pressure which is deposited in the cocoon. In the case of adiabatic jets we find that the the magnetic effects above on the jet structure are stronger.

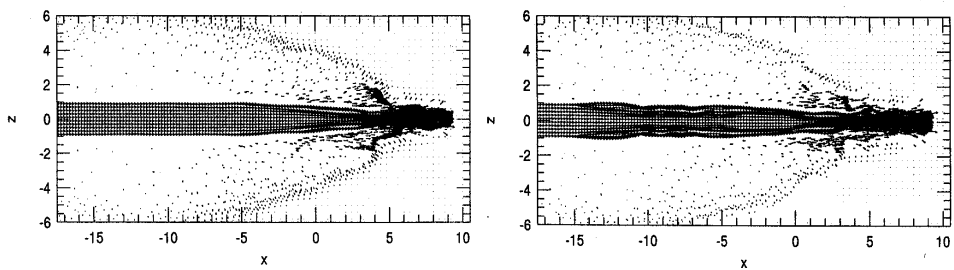

FIGURE 4. Velocity distribution of a radiative cooling HD (left) and a MHD jet (right) with $\beta = 1$ propagating into a homogeneous ambient medium at $t = 1.5$ (in units of 38.2 yr). The initial conditions are the same as in Fig. 2.

5. CONCLUSIONS

We have presented fully 3-D HD and MHD simulations of radiative cooling jets propagating into homogeneous and stratified ambient media in order to clarify the structural properties of YSO jets. The simulations show that the dense, cold shell that develops at the head becomes dynamically unstable and fragments into pieces that resemble the observed HH objects.

The observed internal knots and multiple bow shocks are probably mainly formed by time-variations in the ejection velocity of the jet. However, jets propagating into an ambient medium with increasing density (and pressure) develop a wiggling chain of knots, close to the jet head, which is driven by Kelvin-Helmholtz instabilities. This morphology is similar to some observed systems.

In an ambient medium of decreasing density (and pressure), the beam widens and relaxes, becoming faint. This could explain the invisible jet sections between the parent source and the beam.

YSO jets seem to transfer momentum predominatly through the bow shocks other than through turbulent entrainment. We can then speculate that molecular outflows could be formed from environmental gas swept by the passage of multiple bow shocks in a time-variable jet.

The presence of helical magnetic fields in a cooling jet (with $\beta \simeq 1$) reduces the growth of the density in the shell, increases jet collimation and promotes some beam pinching which may also cause the formation of internal knots along the beam within a few dynamical times. The latter effect, however, is inhibited by the presence of the radiative cooling.

The authors are in debted with the referee Dr. K. Tsinganos for fruitful suggestions on this work. The simulations were performed on the DEC3000/600 AXP workstation whose purchase was made possible by FAPESP. This work was partially supported by CNPq.